\documentclass[
10pt
,aps
,english
,floatfix
,longbibliography
,nofootinbib
,prx
,superscriptaddress
,twocolumn
]{revtex4-2}

\usepackage{amsfonts}
\usepackage[english]{babel}
\usepackage{booktabs}
\usepackage{braket}
\usepackage{csquotes}
\usepackage{enumitem}
\usepackage[T1]{fontenc}
\usepackage{graphicx} 
\usepackage[colorlinks
,breaklinks
,urlcolor={blue}
,linkcolor={red}
,citecolor={blue}]{hyperref}
\usepackage{mathrsfs}
\usepackage{mathtools}
\usepackage[caption=false]{subfig}   
\usepackage{xurl}

\DeclareQuoteStyle{english}%
    {\textquoteleft}
    [\textquoteleft]
    {\textquoteright}
        [0.05em]   
    {\textquotedblleft}
    [\textquotedblleft]
    {\textquotedblright}

\graphicspath{ {images/} }

\begin{document}
\setquotestyle{english}
\title{t\={u}Q: a design and modelling tool for cluster-state algorithms}
\begin{abstract}
This paper is a general introduction to the t\={u}Q toolchain (\url{https://github.com/QSI-BAQS/tuQ}) and a discussion of its two main workflows, \enquote{reduce and optimise} and \enquote{draft and compile}.  The t\={u}Q toolchain was designed to advance research in cluster-state computing and the workflows are presented as suggestions for how a researcher might use the tool.  The two modes of t\={u}Q are Modeller and Simulator.  Simulator mode has a tile-based syntax for drafting cluster-state algorithms.  Modeller enables the user to reduce a lattice through preset measurement functions and optimise an algorithm by minimising the count of qubits or the count of controlled-Z ($CZ$) interactions.  In addition, t\={u}Q makes it possible to compile an algorithm to OpenQASM 3.0.
\end{abstract}
\author{Greg Bowen}
\affiliation{Centre for Quantum Software and Information, University of Technology Sydney, Sydney, NSW 2007, Australia}
\author{Simon Devitt}
\affiliation{Centre for Quantum Software and Information, University of Technology Sydney, Sydney, NSW 2007, Australia}
\maketitle

\section{Introduction}
The open-source application, t\={u}Q is a toolchain for modelling, optimising and compiling an algorithm that is consistent with the cluster-state model of quantum computation.  As compared to the many open-source, circuit-based emulators and compilers currently available\footnote{ An extensive catalogue of emulator libraries or online resources is available at URL: \url{https://qosf.org/project_list/}.}, there are few open-source applications for emulating cluster-state computing or for drafting such algorithms.  Such tools as Graphix \cite{SF22} or McBeth \cite{EOS23} let their user express a cluster-state algorithm, which defines an abstract cluster state as part of its output.  As will be shown, t\={u}Q enables its user both to draft an algorithm and model a cluster state directly, \textit{each as an expression of the other}.  The use case of t\={u}Q therefore includes practical workflow for iterative drafting and refining of a cluster-state algorithm.  The two modes of t\={u}Q are Modeller and Simulator.  Modeller mode is for modelling cluster states through the use of simple graphs, with the aim of minimising the number of qubits required to compute an algorithm.  Simulator mode is for drafting algorithms in a tile-based syntax with established cluster-state patterns and single-qubit measurements for the user to \enquote{roll your own} widgets.  

As a model of quantum computation, a cluster state is a system of \textit{n}-entangled qubits through which an algorithm is processed by a combination of single-qubit measurements and propagation of state by means of quantum teleportation.  A cluster state has a topology of dimensions $\mathbb{Z}^d$, where $d\hspace{1mm}\text{(dimension)} \geq1$ although convention, also observed in this paper, restricts it to a two-dimensional lattice (cf. \cite{BDP16}).  While researchers have proposed more than one cluster-state model (e.g. \cite{CLN05,BR05}), the design principles of t\={u}Q are based on measurement-based quantum computing (\enquote{MQBC}), which is the founding and best known instance of cluster-state computation \cite{RB02,RBB03,RHG06}.  The cluster state of MBQC, denoted $QC_\mathcal{C}$ is referred to as the \enquote{substrate} of computation.  A detailed discussion of $QC_\mathcal{C}$, including its preparation and deterministic computational properties can be found at \cite{RB01,RB02}.

A cluster state is more commonly known as a \textit{graph state} (e.g. \cite{HEB04,Nie06}) and has been quantified as an alternative model of quantum computation \cite{VDD04,HDE06,CDL11,DHW20,VPG24}.  A graph, $G$, is an ordered pair consisting of a non-empty, finite set of vertices, $V\{G\}$ and a finite set of edges, $E\{G\}$.  As a graph \textit{state}, $\ket{G}$, vertices are qubits and edges are interactions within a \textit{simple} graph structure thereby prohibiting both an edge as a \enquote{self-join} of a qubit and more than one edge between vertices.

This paper is an introduction to the t\={u}Q toolchain with a specific focus on using it as part of workflow for designing and optimising of a cluster-state algorithm.  Section \ref{tuQ} is a general introduction to the toolchain, including its use case while Section \ref{workflow} is a demonstration of t\={u}Q's two main workflows, \enquote{reduce and optimise} and \enquote{draft and compile} as suggestions of how the toolchain might be used as part of research into cluster-state computation.

\section{the t\={u}Q toolchain}
\label{tuQ}
A design priority of the t\={u}Q toolchain was to advance research in cluster-state computing.  To that end, the t\={u}Q user can,
\begin{enumerate}
   \item read a circuit encoded in JSON to convert it to a graph state,
   \item draft an algorithm in a tile-based syntax that is consistent with cluster-state principles,
   \item model and reduce a cluster-state to identify possible optimisations of its originating algorithm; and
   \item transpile an algorithm to OpenQASM 3.0.
\end{enumerate}
This paper is a demonstration of t\={u}Q functions 2-4; reading a circuit is simply a different method for obtaining an algorithm to model as $\ket{G}$.
\begin{figure*}
    \includegraphics[width=0.8\linewidth]{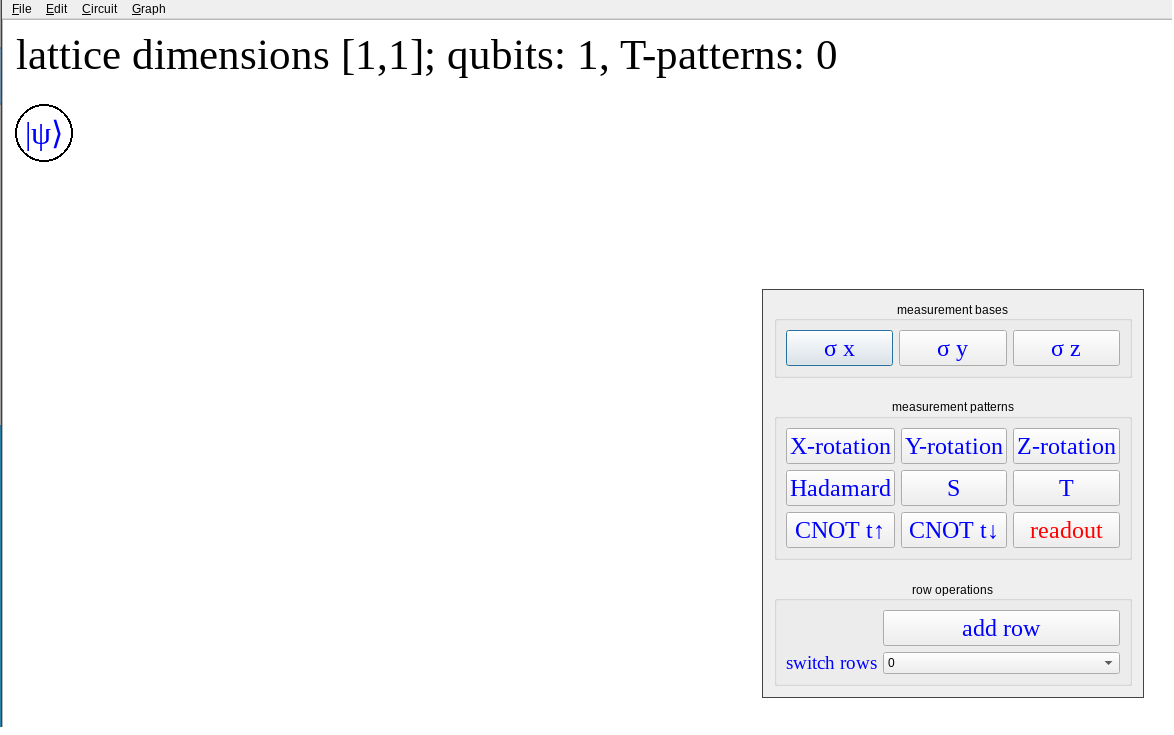}
    \caption{\label{fig:tiles}The tiles palette (bottom, right) from which a user can select standardised measurement patterns to encode its algorithm.  Upon launching t\={u}Q Simulator, the user can access the palette and has the real-time data at the top margin of the canvas, which include from left-to-right: minimum $\ket{G}$ in [row, column], the number of qubits in $\ket{G}$ and a count of $T$ patterns.  These data enable a user to track the size of the lattice while drafting an algorithm; and are used to generate the lattice through the menu function, \enquote{Open Algorithm}.  The count of $T$ patterns is a gauge of the additional qubits required for magic state distillation.}
\end{figure*}

The origin of t\={u}Q Modeller mode is the standalone application for shaping $\ket{G}$, Q2Graph.  An introduction to Q2Graph and relevant functions of the application can be found in \cite{BD22}.  The requirements for t\={u}Q as an application for modelling $\ket{G}$ were,
\begin{enumerate}[label=\roman*]
   \item an interactive tool for drawing a two-dimensional simple graph as a prototype of $\ket{G}$,
   \item functions to reproduce the effect of Pauli group operations on $\ket{G}$; and
   \item demonstrate LC-equivalence of $\ket{G}$ and $\ket{G'}$.
\end{enumerate}
In terms of scale, Modeller can create a lattice of maximum dimensions $\left[121, 121\right]$.  

The precursor to t\={u}Q Simulator was an application with the working title of \enquote{Etch}, which was created for testing a graph-state compilation proposal known as \textit{circuit etching} \cite{BCD25}.  The user of t\={u}Q Simulator can,
\begin{enumerate}
   \item write an algorithm expressly for $\ket{G}$; and
   \item derive the size of the lattice medium dictated by 1., in real time; and
   \item transpile an algorithm to assembly language, OpenQASM 3.0
\end{enumerate}
As such, this t\={u}Q mode partly fulfils the role of both quantum emulator and (compiling) optimiser.  Simulator provides a tile-based syntax for the user to write a graph-state algorithm (see Figure \ref{fig:tiles}).  

\begin{figure*}
    \includegraphics[width=0.9\linewidth]{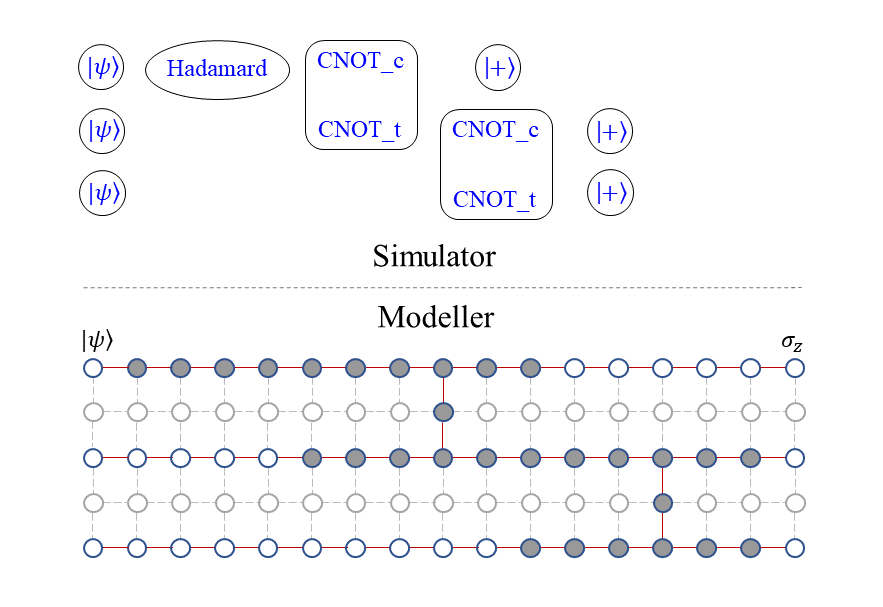}
    \caption{\label{fig:patterns}Simulator tiles map to a specific configuration of Modeller vertices and edges.  In \textit{Simulator} a tile-based format of rounded rectangles represents standardised measurement patterns placed with the tiles palette of Figure \ref{fig:tiles}.  The leftmost column, tagged $\ket\psi$, represents input qubits prepared in an arbitrary state; the rightmost column, tagged $\ket{+}$ , represents the readout columns prepared in state $\frac{\ket{0} +\ket{1}}{\sqrt{2}}$.\\
    In \textit{Modeller} layout of $\ket{G}$, qubits (vertices) appear as circles while an edge between neighbouring qubits represents a controlled-Z ($CZ$) interaction.  The leftmost column, labelled $\ket\psi$, consists of $\ket{G}$ input qubits and the rightmost column, labelled $\sigma_{z}$, consists of readout qubits to be measured in the computational basis.  Qubits with a solid fill indicate basis $\sigma_{x}$/$\sigma_{y}$ measurements while qubits with no fill, excepting the readout qubits, are measurements in basis $\sigma_{x}$.  The fainter, \enquote{greyed-out} qubits are removed from $\ket{G}$ by basis $\sigma_{z}$ measurements ahead of any measuring of computational qubits (see \cite{RB02} for details of preparing a cluster state for computation).}
\end{figure*}

The tiles of Simulator are an approximation of measurement \textit{patterns} after \cite{DKP09}.  Measurement patterns or simply patterns encompass the preparation, measurement and requisite corrections of computational state in cluster-state $\ket{\Phi}$.  A pattern is the common unit between Simulator and Modeller because a Simulator tile represents a standardised sequence of measurements that also maps to a specific configuration of Modeller vertices and edges (see Figure \ref{fig:patterns}).  For example, the Simulator tiles for the Clifford group operators, Hadamard, $S$ and CNOT each correspond to a standardised configuration of vertices, edges and single-qubit measurements as defined in \cite{RB02}.  Simulator adds the non-Clifford $T$ rotation to its tiles palette to provide a universal quantum gate set \cite{BBC95} and uses the general rotation pattern, $U_{Rot}$ of \cite{RB02} as a proxy for the lattice layout of $T$. 

If the user wishes to compute an algorithm drafted in t\={u}Q then, Simulator mode includes a function to transpile a cluster-state algorithm to the open-source quantum assembly language, OpenQASM 3.0 \cite{CJA22}.  Classical assembly language is a low-level syntax, which translates directly to the machine code compatible with a processor.  Many providers of NISQ (\enquote{Noisy intermediate-scale quantum}) computing services use the quantum equivalent of classical assembly language, usually OpenQASM, as an additional medium for passing algorithms to their quantum processors via an API.  The aim for t\={u}Q is thus to render an algorithm drafted for $\ket{G}$ as computable on a QPU configured for OpenQASM.

\section{workflow}
\label{workflow}
There are two main branches of workflow available through t\={u}Q.  The user can,
\begin{itemize}
    \item \enquote{reduce and optimise},
    \item \enquote{draft and compile}.
\end{itemize}
Each workflow branch is addressed below.

\subsection*{reduce and optimise}
The t\={u}Q user can transfer its algorithm drafted in Simulator mode directly to Modeller mode in which it can then manually reduce $\ket{G}$ as a replica of operations within a quantum processor.  This workflow branch is usually iterative and allows the user to reason both about propagation of quantum information through the lattice and consequently, how the algorithm might be optimised to shrink the lattice.  Moreover, the user has the option at any point in the process to compute its algorithm by means of encoding it to OpenQASM 3.0 (see \enquote{draft and compile}).

\begin{figure*}
    \includegraphics[width=\linewidth]{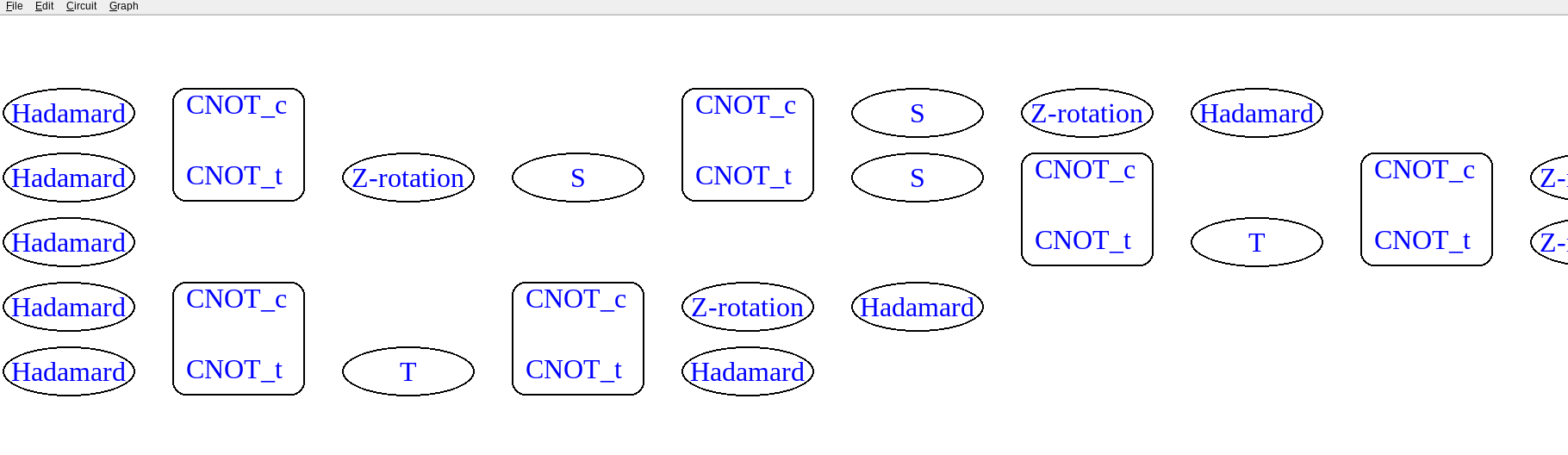}
    \caption{\label{fig:refit}Reformatting the IQP circuit input directly using Simulator's tile syntax.}
\end{figure*}
\begin{figure*}%
   \subfloat[\label{fig:oa}The \enquote{Open Algorithm} function lays out patterns or basis $\sigma_{x}/\sigma_{y}/\sigma_{z}$ measurement qubits within the lattice, highlights patterns as Clifford (red perimeter) or non-Clifford (blue perimeter) and displays the component measurements of a pattern.  Vertices marked with a (qubit) integer ID require further processing.]{
   \includegraphics[width=0.75\linewidth]{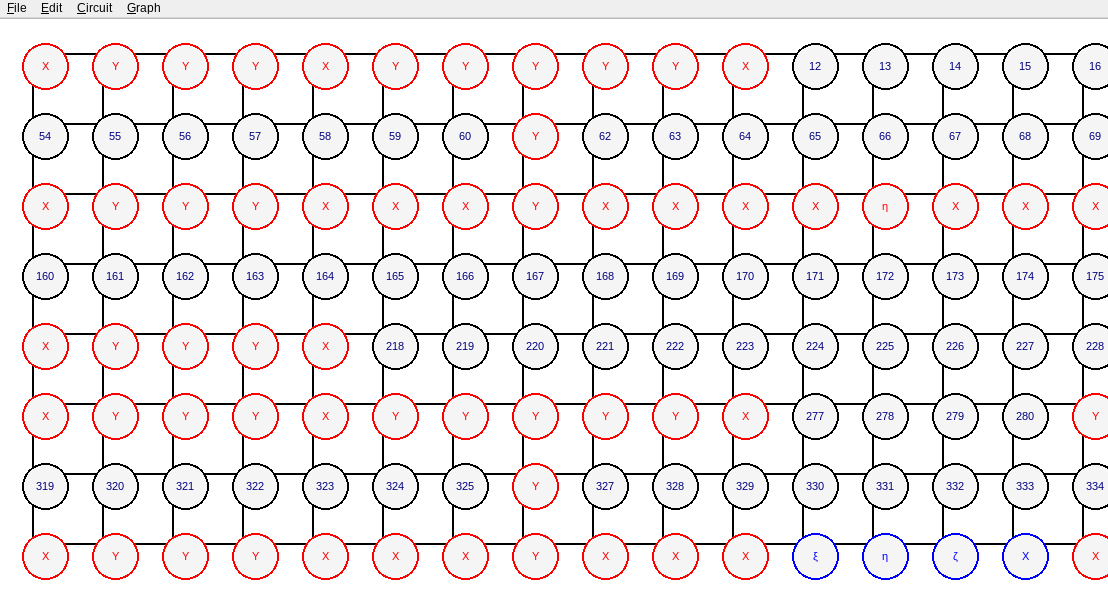}
   }\\
   \subfloat[\label{fig:oaz}Manual steps to follow the Open Algorithm function include removing superfluous qubits of the lattice by measuring in basis $\sigma_{z}$.  Any \enquote{standard} qubit that precedes or follows a qubit highlighted by the function is to be measured in basis $\sigma_{x}$ as part of computing through the lattice.  This operation replicates the left-to-right teleporting of state through the lattice.]{
   \includegraphics[width=0.75\linewidth]{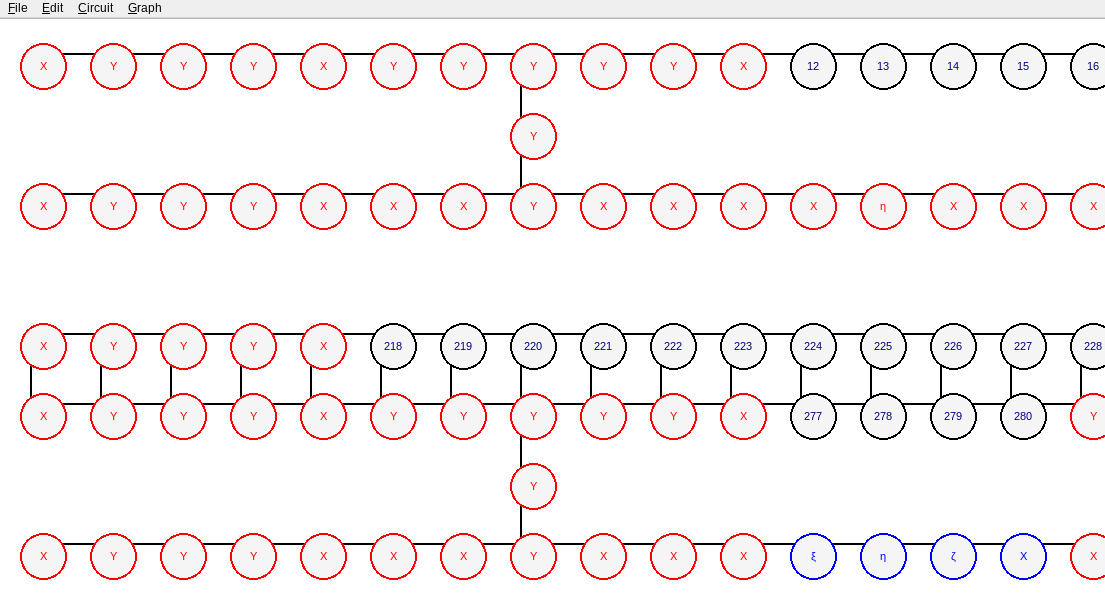}
   }%
   
   \caption{\label{fig:oaOaz}Western extremity of the lattice specified by the [8,53] algorithm and rendered by t\={u}Q Modeller function, \enquote{Open Algorithm}, showing automated and manual steps of preparation.}   
\end{figure*}

Consider Figure \ref{fig:refit}, which is a representation of the western extremity of an algorithm drafted through Simulator\footnote{ See Appendix \ref{demo} for [row, column] coordinates of all patterns of this algorithm.}.  The algorithm specifies an [8,53] lattice, which is unwieldy both for drawing by hand and subsequently for locating the individual patterns within it.  In a case such as this, the user has the option of the Modeller menu function, \enquote{Open Algorithm} which plots the algorithm as a lattice.  The steps of the Open Algorithm function include,
\begin{enumerate}[label=\roman*.]
   \item creating the minimum required lattice, then
   \item situating the pattern or basis $\sigma_{x}/\sigma_{y}/\sigma_{z}$ measurement qubits within the lattice; and
   \item highlighting each Clifford pattern or basis $\sigma_{x}/\sigma_{y}/\sigma_{z}$ measurement qubits in \textit{red}.  The single non-Clifford $T$-pattern highlights in \textit{blue}; and
   \item specifying each measurement of the pattern that is needed to replicate the equivalent (circuit) gate, in place of a vertex (qubit) integer ID.  Rotation patterns along the X/Y/Z axis carry the relevant $\xi, \eta, \zeta$ Euler denotation while the \enquote{general rotation} pattern as supplied by Rau{\ss}endorf and Briegel \cite{RB02} stands in for the non-Clifford $T$-pattern.
\end{enumerate}
Figure \ref{fig:oa} is a demonstration of the [8,53] algorithm of Figure \ref{fig:refit} as rendered through the Open Algorithm function.  Vertices marked with a (qubit) integer ID require further processing.  In order,
\begin{enumerate}
    \item measure superfluous qubits in basis $\sigma_{z}$ to remove them from the lattice (Figure \ref{fig:oaz}); then,
    \item Any remaining integer ID qubit (e.g. Figure \ref{fig:oaz}, qubits 12-16) is to be measured in basis $\sigma_{x}$ to use the qubit(s) as a \enquote{wire}.  A wire effectively propagates state by teleportation, left-to-right through the lattice.
\end{enumerate}

The user can now reduce the lattice.  It is expected that in most instances a user will reduce the lattice as guided by the measurement label of a given vertex.  In doing so, the user will actively look for opportunities to minimise the lattice to an equivalent lattice without information loss in much the same way as classical intermediate representations work as a device of iterative optimisation.  This strategy hinges upon the property of local complementation (\enquote{LC}) equivalence.

The complement of graph $G$ is $G'$, which has the same set of vertices, $V(G)$ but inverted edges.  \textit{Local} complementation restricts the complement to a neighbourhood of $G$, leaving the rest of $G$ unaffected.  In regards to $\ket{G}$, iff $\ket{G'}$ is $\ket{G}$ following at least one local complementation operation, $\ket{G}$ and $\ket{G'}$ are termed LC-equivalent \cite{HDE06}.  Note, local complementation is a component of the preset $X/Y/Z$ Pauli operations of t\={u}Q Modeller.  In t\={u}Q Modeller, after Hein \textit{et al}. \cite{HDE06}, measuring the qubit represented by vertex $a$ in basis $\sigma_{x}$, $\sigma_{y}$ or $\sigma_{z}$ will transform $\ket{G}$ to $\ket{G'}$ on the remaining vertices thus,
\begin{align*}
   Z &: P^{a}_{z,\pm}\ket{G} = \frac{1}{\sqrt{2}}\ket{z,\pm}^{a} \otimes U^{a}_{z,\pm}\ket{G - a},\\
   Y &: P^{a}_{y,\pm}\ket{G} = \frac{1}{\sqrt{2}}\ket{y,\pm}^{a} \otimes U^{a}_{y,\pm}\ket{\tau_{a}(G) - a},\\
   X &: P^{a}_{x,\pm}\ket{G} = \frac{1}{\sqrt{2}}\ket{x,\pm}^{a} \otimes U^{a}_{x,\pm} \ket{\tau_{b_{0}} (\tau_{a} \circ \tau_{b_{0}}(G) - a)},\\    
\end{align*}
where $P^{a}$ signifies the projection of vertex $a$ into the designated $_{x/y/z}$ basis and $\tau_{a}$ denotes local complementation of vertex $a$'s neighbourhood, which is the subgraph of all vertices linked by an edge to vertex $a$.  Again, after \cite{HDE06} and in plain English in which $N_a$ denotes the neighbourhood of vertex $a$,
\begin{itemize}
    \item $Z$: delete vertex $a$ from $G$,
    \item $Y$: invert $G \left[ N_a\right]$ and delete vertex $a$ from $G$,
    \item $X$: choosing any $b_0 \in N_a$, invert $G \left[ N_{b_0}\right]$, applying the rule for $Y$ then, invert $\tilde{G} \left[ N_{b_0}\right]$ again.
\end{itemize}
Importantly, LC-equivalence in a stabiliser state, to which $\ket{G}$ qualifies \cite{AB06,HDE06}, ensures continuity of state-space through successive measurements.

Alternatively, the user could use the local complementation function (Modeller: keystroke \enquote{o}) directly at any point to test for LC-equivalence as a possible optimisation.  As shown, Pauli measurements through t\={u}Q Modeller delete both vertices and edges from $\ket{G}$ whereas local complementation inverts edges and does not affect the number of vertices.  Any optimisation realised through LC-equivalence manifests as a reduction in the number of controlled-Z ($CZ$) gates required to create $\ket{G}$.  A $CZ$ gate on vertices of $\ket{G}$ is a two-qubits interaction such that,
\[
   CZ = diag(1,1,1,-1),
\]
in which one qubit acts as a control on the state of the other target qubit.  Canonically, $CZ$ effects the entanglement correlation between qubits as a requirement of a cluster state \cite{CDL11,DHW20}.  Minimising the count of $CZ$ interactions required to create a cluster state is as much an optimisation of $\ket{G}$ as minimising the count of its qubits.

\begin{figure*}
    \includegraphics[width=0.8\linewidth]{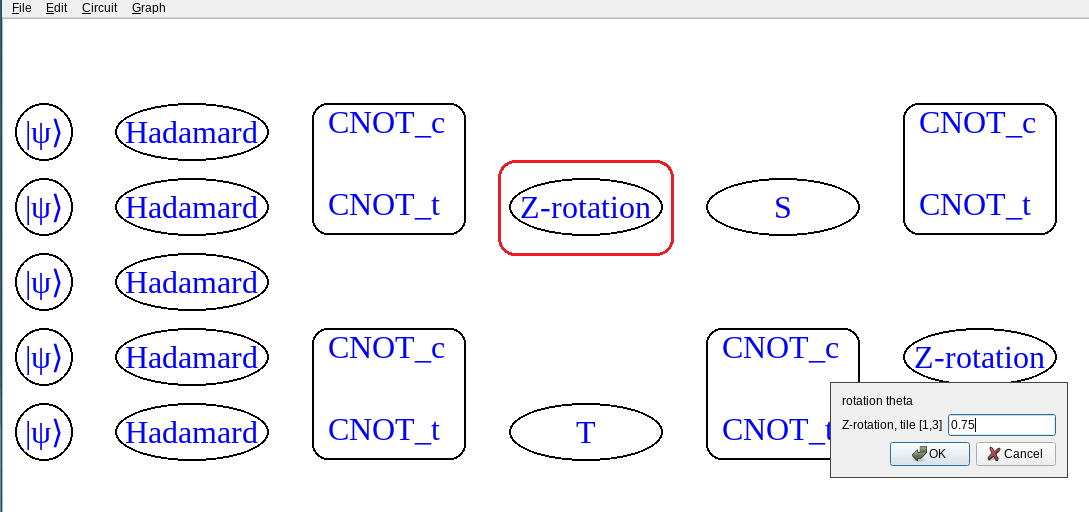}
    \caption{Direct input of $\theta$ as part of the t\={u}Q Simulator function, \enquote{Compile}.  In this example, the input dialog matches the Z-rotation tile at position [1, 3], as outlined in red, with an input field for the user to set the requisite $\theta$.}
    \label{fig:theta}
\end{figure*}

\subsection*{draft and compile}
The user of t\={u}Q may bypass any \enquote{reduce and optimise} workflow, preferring instead to compute its algorithm through a quantum computing service.  The Simulator mode's menu function, \enquote{Compile} is central to this workflow branch.

Assuming t\={u}Q is suitably configured for it\footnote{ The compile component of t\={u}Q requires some customising at load: the user must specify the API URL as variable, reqStrQASM.}, the Compile function will,
\begin{enumerate}
   \item encode the base algorithm to OpenQASM 3.0 and a copy of the script will write to a .txt file for inspection.  This is the default setting of the Compile function; and
   \item POST the script to the supplied (REST) API.  With reference to this function, the API must conform with OpenQASM 3.0 syntax for POST to succeed.  In such cases that OpenQASm 3.0 is incompatible with a quantum computing API, it is of course possible to transcribe 1. to an earlier version of OpenQASM then POST it manually.
\end{enumerate}
Appendix \ref{openQASM} is the OpenQASM 3.0 transcription of the cluster-state algorithm.  The user must specify $\theta$ of any X/Y/Z rotation patterns transcribed to OpenQAS as rotation gates rx/ry/rz.  The Compile function enables the user to supply the rotation $\theta$ through a dialog box, as shown in Figure \ref{fig:theta}.  Note, text specifying the rotation pattern and its [row, column] coordinates within the lattice appears in the Compile dialog to remove ambiguity that might otherwise arise when transpiling a large algorithm with multiple rotations.

Clearly OpenQASM is intended for a gate-based processor which prompts the question, is t\={u}Q actually compiling a cluster-state algorithm?  Transpiling a cluster-state algorithm to a circuit-based processor at the very least carries a risk of information loss from the algorithm for example, cluster states readily admit composition of individual measurements into patterns whereas quantum gates do not compose (cf. \cite{GGW24}).  OpenQASM necessarily imposes circuit-model algorithms on a cluster-state toolchain.  A broader view is to subordinate the discrepancy between graph-state algorithm and circuit-based processor to the outcome of an algorithm obtaining meaningful output \textit{within these restrictions}.  Furthermore, an assembly language organised around the cluster-state model may not exist as at December 2024 but should that variant eventuate, t\={u}Q's Compile function can refactor from OpenQASM to it.  In other words, t\={u}Q paired with OpenQASM is proof-of-concept compiling that showcases output from a graph-state algorithm.

\section{discussion}
This paper serves as both a general introduction to the t\={u}Q toolchain and a specific discussion of its two main workflows.  The toolchain is suitable for modelling, optimising and compiling a cluster-state algorithm.  Comparatively few open-source applications allow a user to draft a cluster-state algorithm and emulate its computation but t\={u}Q facilitates this combination of functionality so that an algorithm and cluster-state is each a direct expression of the other.  Modeller mode is for modelling cluster states through the use of simple graphs, with the aim of minimising the number of qubits required to compute an algorithm.  The tile-based syntax of Simulator mode is for drafting algorithms with preset patterns or single-qubit measurements.

The t\={u}Q toolchain was designed to advance research in cluster-state computing.  The two main workflows, \enquote{reduce and optimise} and \enquote{draft and compile} are presented as suggestions of how the toolchain might be used as a research tool.  Modeller's preset measurement functions and its local complementation function mean the user can optimise its algorithm by minimising the count of qubits, or the count of $CZ$ interactions required to resource a cluster state.  Moreover, t\={u}Q makes it possible to compile an algorithm to OpenQASM 3.0 as a proxy for quantum computing services that do not natively use graphs.

The final consideration for the toolchain and indeed, quantum compilers in general, is compensating for the probabilistic nature of measuring quantum state.  With regard to the MBQC framework, applying extra rotations in basis $\sigma_{x}$ or $\sigma_{z}$ to the relevant readout qubit is sufficient to correct $\ket{G}$ for randomness of measurements \cite{RB02}.  To effect the correcting $\sigma_{x}/\sigma_{z}$ rotations to readout qubits requires a log of measurements of every qubit in the system.  This log, known alternately as a pauli- \textit{tracker} \cite{PDN14} or \textit{frame} \cite{RFV17}, comprises of (pauli) records of every measurement of any qubit to teleport state to the readout qubit under consideration.  A pauli tracker library, compatible with the cluster-state computing model, is available \cite{RD24} although it is not included in the version of t\={u}Q introduced in this paper.  This is in large part because of the requirements attaching to this version of t\={u}Q but is also a result of t\={u}Q's Compile function being paired with OpenQASM and by extension, a gates-based processor that essentially makes pauli tracking redundant.

\begin{acknowledgments}
The views, opinions and/or findings expressed are those of the author(s) and should not be interpreted as representing the official views or policies of the Department of Defense or the U.S. Government.  This research was developed with funding from the Defense Advanced Research Projects Agency [under the Quantum Benchmarking (QB) program under award no. HR00112230007 and HR001121S0026 contracts].
\end{acknowledgments}

\nocite{*}

\bibliographystyle{apsrev4-2}
\bibliography{references}

\clearpage
\appendix
\section{\label{demo}coordinates of cluster-state algorithm}
Coordinates are [row, column] and zero-based.  CNOT complies with Rau{\ss}endorf and Briegel's \cite{RB02} 15-qubits pattern therefore \enquote{row} is the southern row of the pattern while \enquote{column} is the eastern edge of the pattern.  Note, horizontally adjacent patterns have their readout and input qubits combined through the property of composition.

\begin{table}[h]
   \begin{center}
      \begin{tabular}{l|r}
         \toprule
         \textbf{pattern} & \textbf{coordinates}\\
         \midrule
         Hadamard & [0, 4]\\
         Hadamard & [2, 4]\\
         Hadamard & [4, 4]\\
         Hadamard & [5, 4]\\
         Hadamard & [7, 4]\\
         CNOT & [2, 10]\\
         Z-rotation & [2, 14]\\
         $S$ & [0, 28]\\
         $S$ & [2, 28]\\
         CNOT & [4, 34]\\
         $T$ & [4, 38]\\
         CNOT & [4, 44]\\
         CNOT & [7, 10]\\
         $T$ & [7, 14]\\
         CNOT & [7, 24]\\
         Z-rotation & [0, 48]\\
         Z-rotation & [2, 48]\\
         Z-rotation & [4, 48]\\
         Z-rotation & [5, 48]\\
         Hadamard & [0, 53]\\
         Hadamard & [2, 53]\\
         Hadamard & [4, 53]\\
         Hadamard & [5, 53]\\
         Hadamard & [7, 53]\\
         \bottomrule
      \end{tabular}
   \end{center}
\end{table}

\section{\label{openQASM}the cluster-state algorithm in OpenQASM 3.0 format}
\includegraphics[width=0.35\textwidth]{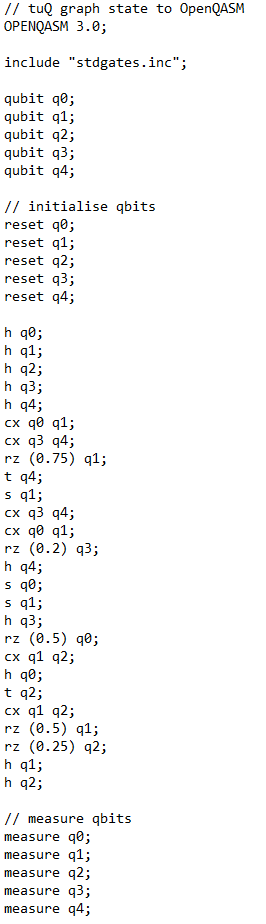}

\end{document}